\documentclass[aps,pre,showpacs,preprintnumbers,amsmath,amssymb,floatfix]{revtex4}
\usepackage{amsmath,graphicx,amsfonts}
\usepackage{dcolumn}
\usepackage{epsfig}
\usepackage{graphicx}
\usepackage{subfigure}

\newcommand{\be}{\begin{equation}}
\newcommand{\ph}{\vec{\phi}} 
\newcommand{\ee}{\end{equation}}

\begin{document}

\title{\bf Scattering of Sine-Gordon kinks on potential wells}

\author{Bernard Piette and W.J. Zakrzewski} 
\affiliation{Department of Mathematical Sciences, University of Durham,
 Science Laboratories, South Road, Durham DH1 3LE, England}

\date{\today}

\begin{abstract}

We study the scattering properties of Sine Gordon kinks on
obstructions in the form of finite size potential `wells'. We model this by 
making the coefficient of the $\cos(\varphi)-1$ term in the
Lagrangian position dependent.
We show that when the kinks find themselves in the well they radiate
and then interact with this radiation. As a result of this energy loss
 the kinks become trapped for small velocities
while at higher velocities they are transmitted with a loss of energy.
However, the interaction with the radiation can produce `unexpected' 
reflections by the well.

We present two simple models which capture the gross features of this 
behaviour. Both involve standing waves either at the edges of the well or in 
the well itself.

\end{abstract}

\pacs{11.10.Lm,12.39.Dc,03.75.Lm}

\maketitle


\section{Introduction}

Recently \cite{one} we have performed a detailed
study of scattering properties
of (2+1) dimensional topological solitons on potential wells.
This work was based on the `baby' Skyrme model, {\it ie} we used
the Lagrangian density which consisted of  three terms:
the pure ${\cal S}\sp2$ sigma model, the Skyrme and the potential terms:
\begin{equation}\label{lagrangian} 
{\cal L}=\partial_{\mu}\ph\cdot\partial^{\mu}\ph
-\theta_{S}\left[(\partial_{\mu}\vec{\phi} \cdot
\partial^{\mu}\vec{\phi})^{2} -(\partial_{\mu}\vec{\phi} \cdot
\partial_{\nu}\vec{\phi}) (\partial^{\mu}\vec{\phi} \cdot
\partial^{\nu}\vec{\phi})\right] -  V(\ph)
\end{equation}
where
\be
V(\phi) = \mu(1-\phi_3^2).
\ee 
The vector $\ph$ was restricted to lie on a unit sphere ${\cal S}^{2}$
hence we put $\ph\cdot\ph=1$.

To generate the potential well the coefficient of the potential term $\mu$ was 
made $x$ dependent. For $x$ outside the well it had one value, say, $\mu_{out}$
and inside the well, {\it ie} for $a<x<b$ its value was reduced to $\mu_{in}$.
This choice of $\mu$ did not affect the vacuum (taken as $\phi_3=+1$); the 
skyrmion was given by a field configuration which varied from $\phi_3=+1$ far 
away from the 
position of the skyrmion to $\phi_3=-1$ at its position). Initially, the 
skyrmions
were placed far away from the `well' (so that all the variation of $\phi_3$ 
from +1 took place for $x$ well away from the `well', {\it ie} for $x<a$).

The `skyrmions' were then sent towards the well and their properties were 
studied.
The obtained results have shown that when the solitons fall into the `well' 
they
get deformed and this deformation may excite the vibrational modes of the
skyrmion and may lead to the skyrmions radiating away their access of energy.
In consequence, the skyrmions can get `trapped' in the well or emerge
from it with a reduced velocity. In \cite{two} we presented a simple four mass 
model
which apes the vibrational modes of the skyrmion and we showed that many of 
the observed
scattering properties of skyrmions can be reproduced in this model - 
suggesting that
their origin resides in the excitation of the lowest vibrational modes of the 
skyrmion.

Given this, it is important to check what happens in models in which solitons
have fewer vibrational modes and so we have decided to look at the (1+1) 
dimensional
Sine-Gordon model and see what happens when its kinks scatter on the 
potential wells.

In the next section we discuss our results obtained for the Sine-Gordon model.
In this model we include a finite size, finite depth, potential well
which is introduced by appropriately modifying the coefficient
of the nonlinear term in the Lagrangian.
The results are qualitatively similar to what we have seen in the 
two-dimentional model and are not very different from the results obtained 
some time ago  by Fei et al 
\cite{older}, in a work which involved the scattering of Sine-Gordon kinks on 
a one-point impurity. 

Recently, Goodman and collaborators \cite{Goodman} have explained these old
results \cite{older} in a `two bounce' resonance model. Their explanation is 
based
on the interaction of the kink with the oscillation of the vacuum (around
the impurity). Thus their model involves two degrees of freedom - the position 
of the kink $x_0$ and the amplitude of the vacuum oscillation
(at the impurity point) $a$. The model of Goodman et al has reproduced all the 
features 
of the results of the original simulations reported in \cite{older}.
Hence in the following section we introduce a similar model for our case
which now involves a finite well of width $2p$ and depth $1-\lambda$.
To do this we make an ansatz for an approximate
field which describes the system. It involves a sine-Gordon kink 
which is allowed to alter its slope and we add to it 
two amplitudes of oscillation of the vacuum (at each end of the `well').
In section 3 we derive the Lagrangian for such an effective model from the 
original Lagrangian.
As the model is somewhat crude we make some drastic approximations in our 
derivation 
of the Lagrangian but still find that the model reproduces the main 
features of the scattering reasonably well. Hence we believe the ideas of 
Goodman and collaborators
to be correct and be more general in nature - thus showing that due to the 
interaction of the soliton with the radiation in the well its behaviour can be
quite complicated and can result in the reflection of the soliton by the well;
{\it ie} a process which is purely classical in nature but could be confused 
with a quantum behaviour. To confirm this further we introduce a further model
(with a couple of radiation standing waves in the well) and again reproduce 
the main features of the full scattering process.

The last section presents some concluding remarks.           

\section{Sine Gordon model and its kinks.}

We take the Lagrangian of the (1+1) dimensional Sine-Gordon model in the form
\begin{equation}\label{lagrangians} 
{\cal L}=\partial_{\mu}\varphi\cdot\partial^{\mu}\varphi
-\lambda^2 \sin(\varphi)^2,
\end{equation}
where, for the kink, the basic field $\varphi$ goes from
$0$ at $x=-\infty$ to $\pi$ at $=\infty$. In the static case its explicit
form is
\begin{equation}
\label{kink}
\varphi(x)\,=\,2 \tan^{-1}(\exp(\theta (x-x_0)),
\end{equation}
where $x_0$ is the kink's position and $\theta$ is its slope. For (\ref{kink})
to be a solution of the equations of motion which follow from 
(\ref{lagrangians}) we need to set $\theta=\lambda$.
 
To have a `well' we set 
\begin{equation}
\label{lambda}
\lambda\,=\,\begin{cases}1\quad \hbox{\phantom{a}for}\quad \vert x\vert > p\cr
\lambda_0\quad \hbox{for}\quad \vert x\vert < p.\cr \end{cases}
\end{equation}
Clearly, $\lambda_0<1$ describes a well while $\lambda_0>1$ 
describes a barrier. As the two-dimensional studies gave more
interesting results for the wells in this paper we restrict our
attention to $\lambda_0<1$.

We have performed many numerical simulations, varying $\lambda_0$ and $p$ (the 
width of the well).

\subsection{Numerical Simulations}

We have performed most of our simulations using a 10001 point lattice with 
the  lattice 
spacing being 0.05.  Hence our lattice extended from -50 to +50.
The kink was initially placed at $x_0=-40$. Its size was determined by 
$\theta=1$
which means that its field was essentially $\varphi\sim 0$ for $x<-45$
and $\varphi\sim\pi$ for $x>-35$.
Thus, there were no problems with any boundary effects (we have verified this
by altering the lattice size and $x_0$).

We have performed three sets of simulations; one involving a very narrow
`well' ($p=0.5$) and two involving a larger well ($p=5$) (one shallow - 
$\lambda_0=0.8$ and one rather deep - $\lambda_0=0.2$). In each case we sent 
the kink
(originally at $x_0=-40$) towards the well varying its initial velocity.
To do this we took the expression for a kink moving with velocity $v$
{\it ie}
\begin{equation}
\label{movingkink}
\varphi(x,t)\,=\,2 \tan^{-1}(\exp(\gamma(x-x_0-vt)),\qquad 
\gamma=\frac{1}{\sqrt{1-v^2}}
\end{equation}
(obtained by Lorentz boosting (4) and setting $\theta=1$) 
and then used it to calculate the initial conditions ($\varphi(x,0)$ and 
${\partial \varphi(x,0)\over \partial  t}$).

All our simulations were performed using a 4th order Runge-Kutta method
for simulating the time evolution.
 The time step of our
simulations was taken to be 0.0001. 
We used fixed boundary conditions and later we used also absorbing boundary 
conditions at the edges of the lattice. This was generated by successively
decreasing the magnitude of ${\partial\varphi\over \partial t}$ for the last 
50 points at both ends of the lattice.

\subsection{Deep Well}

In the deep well case we took $\lambda_0=0.2$ and $p=5$. We have found that 
when the kink was in the well it radiated
and so, when it finally emerged from the well its velocity was lower than the 
initial velocity. This was due to the fact that the well distorted the soliton
which then began vibrating ({\it ie} its slope started oscillating).
These vibrations were then gradually converted into radiation with the slope
settling at its original value.
The curve of the outgoing velocity as a function of the incoming one is shown 
in fig. 1.
\begin{figure}
\begin{center}
\includegraphics[width=10cm]{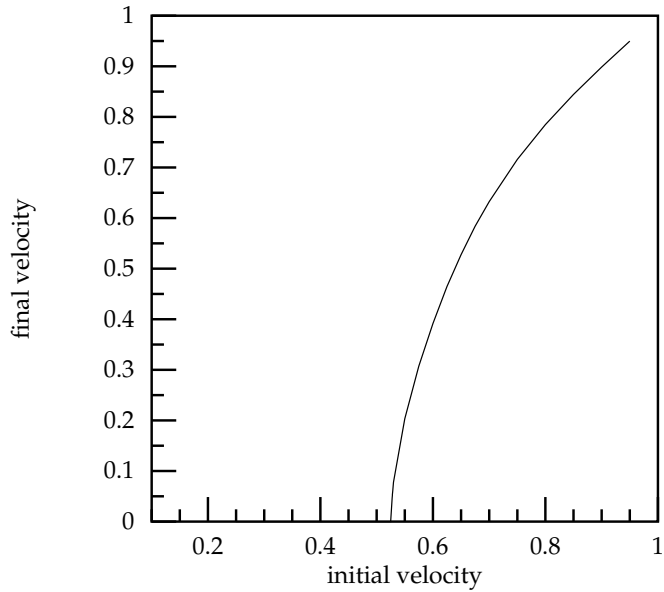}
\caption{Velocity of the kink leaving the well as a function of  
its velocity as it approaches the well. $p=5$, $\lambda_0=0.2$.}
\end{center}
\end{figure}
As is clear from the plot, the kink whose initial velocity is less than 
$v_{cr}\sim  0.527$ gets 
trapped in the well. The curve is very smooth and, as expected, we note that 
for incoming velocities larger than $v_{cr}$ the outgoing velocity is always 
larger than the incoming one demonstrating the loss of the kinetic energy
of the kink through vibration resulting in radiation.

\subsection{Shallow Well}

For a shallow well we took $\lambda_0=0.8$ and still $p=5$. This time the 
critical velocity is much smaller
(as the well perturbs the kink much less).
The curve of the outgoing velocity as a function of the incoming one is shown 
in fig. 2a.

\begin{figure}[htbp]
\unitlength1cm \hfil
\begin{picture}(16,8)
 \epsfxsize=8cm \put(0,0){\epsffile{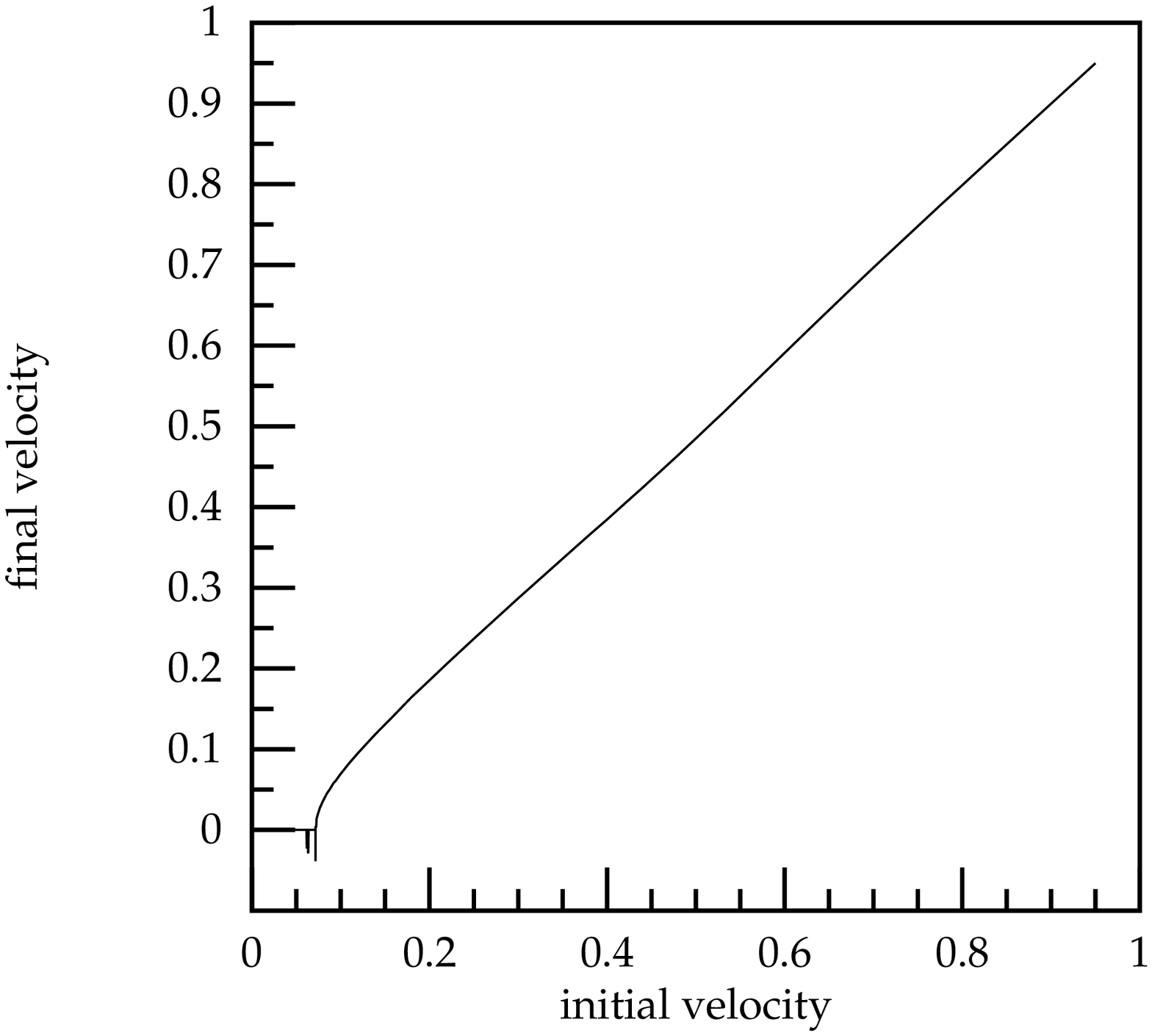}}
 \epsfxsize=8cm \put(8,0){\epsffile{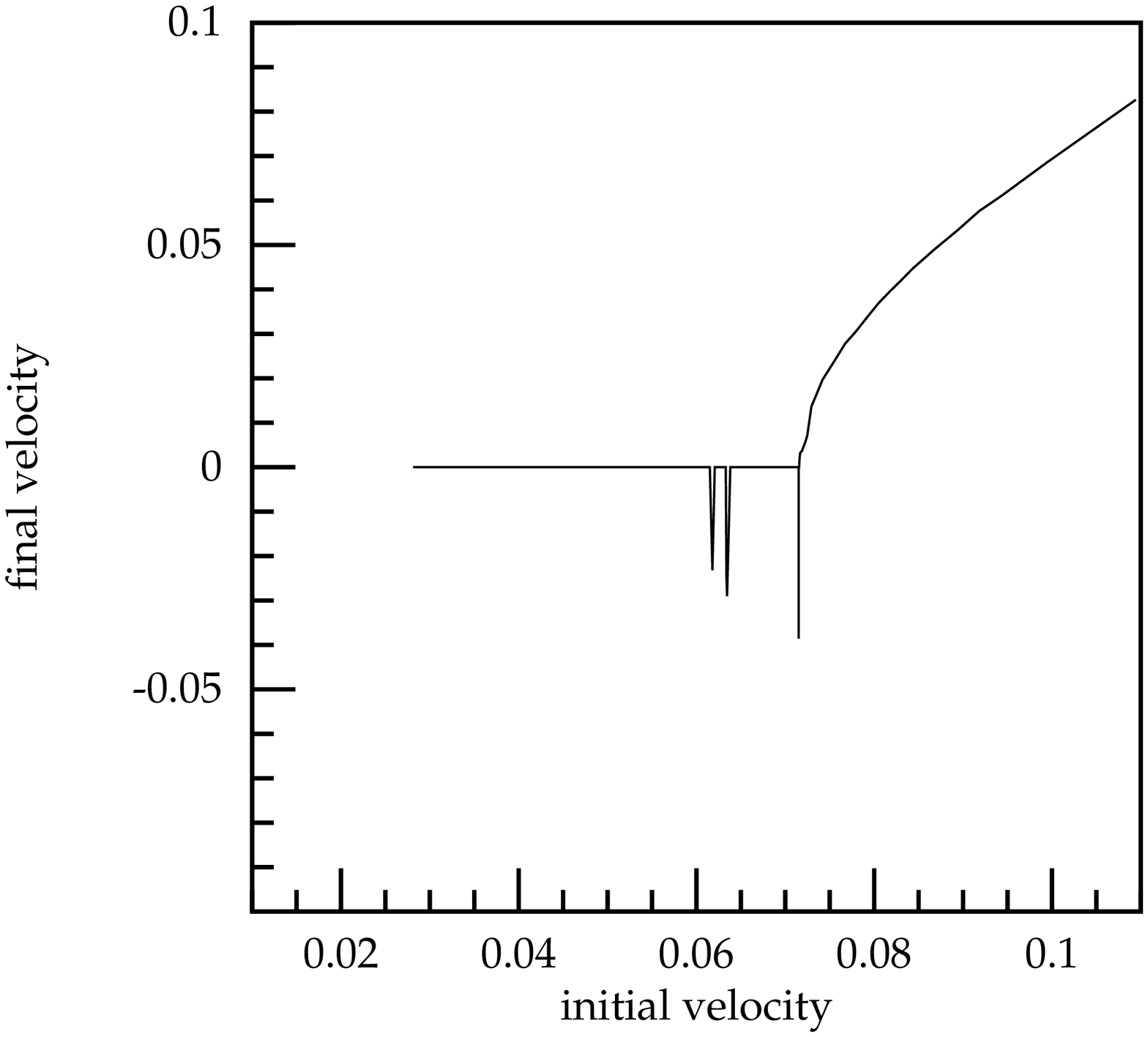}}
\put(4,0){a}
\put(12,0){b}
\end{picture}
\caption{\label{vout_L10k} Velocity of the kink leaving the well as a 
function of  
its velocity as it approaches the well ($p=5$, $\lambda_0=0.8$). a) Full plot. 
b) A close up for small velocities.}
\end{figure}

Looking at the plot we note that there is some irregularity close to the 
threshold. Blowing it up (see fig. 2b) 

we note that just below the critical velocity 
we also have some negative velocities ({\it ie} the whole process looks as if 
the kink was reflected by the well!).
Thus in addition to trajectories like those in fig. 3a and fig. 3b we also 
 have trajectories  like  those of fig. 3c

\begin{figure}[htbp]
\unitlength1cm \hfil
\begin{picture}(16,16)
 \epsfxsize=8cm \put(0,8){\epsffile{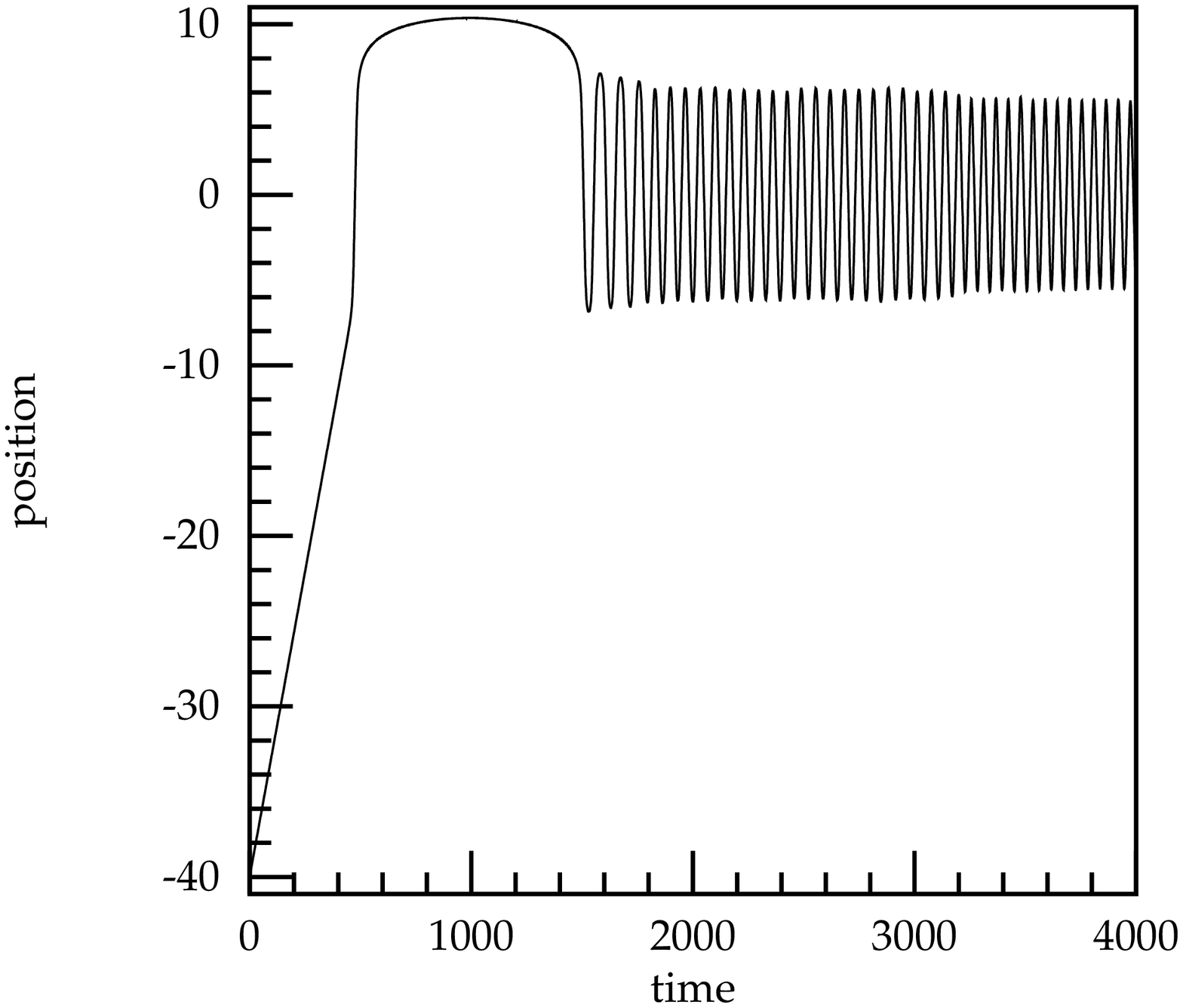}}
 \epsfxsize=8cm \put(8,8){\epsffile{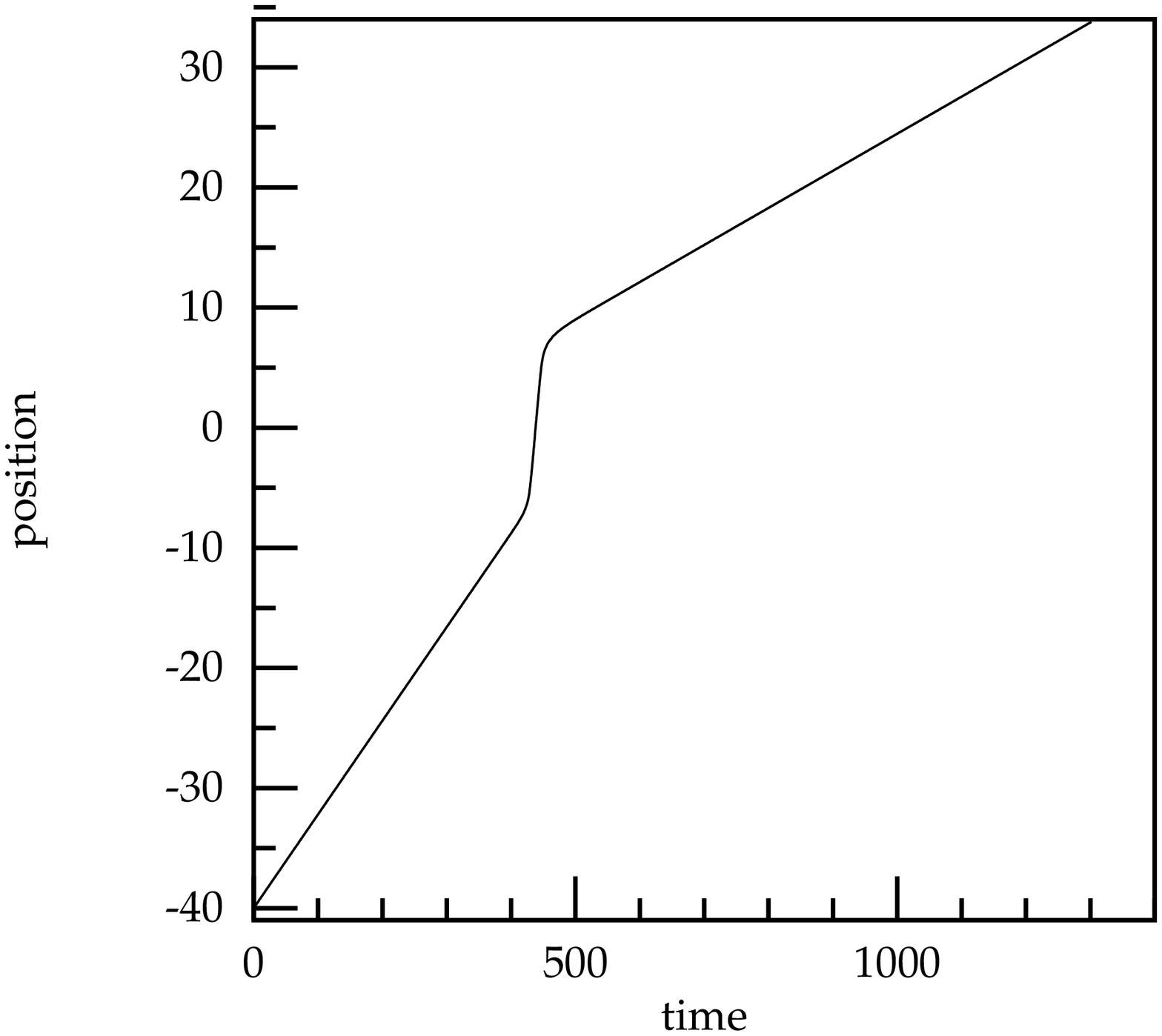}}
\epsfxsize=8cm \put(4,0){\epsffile{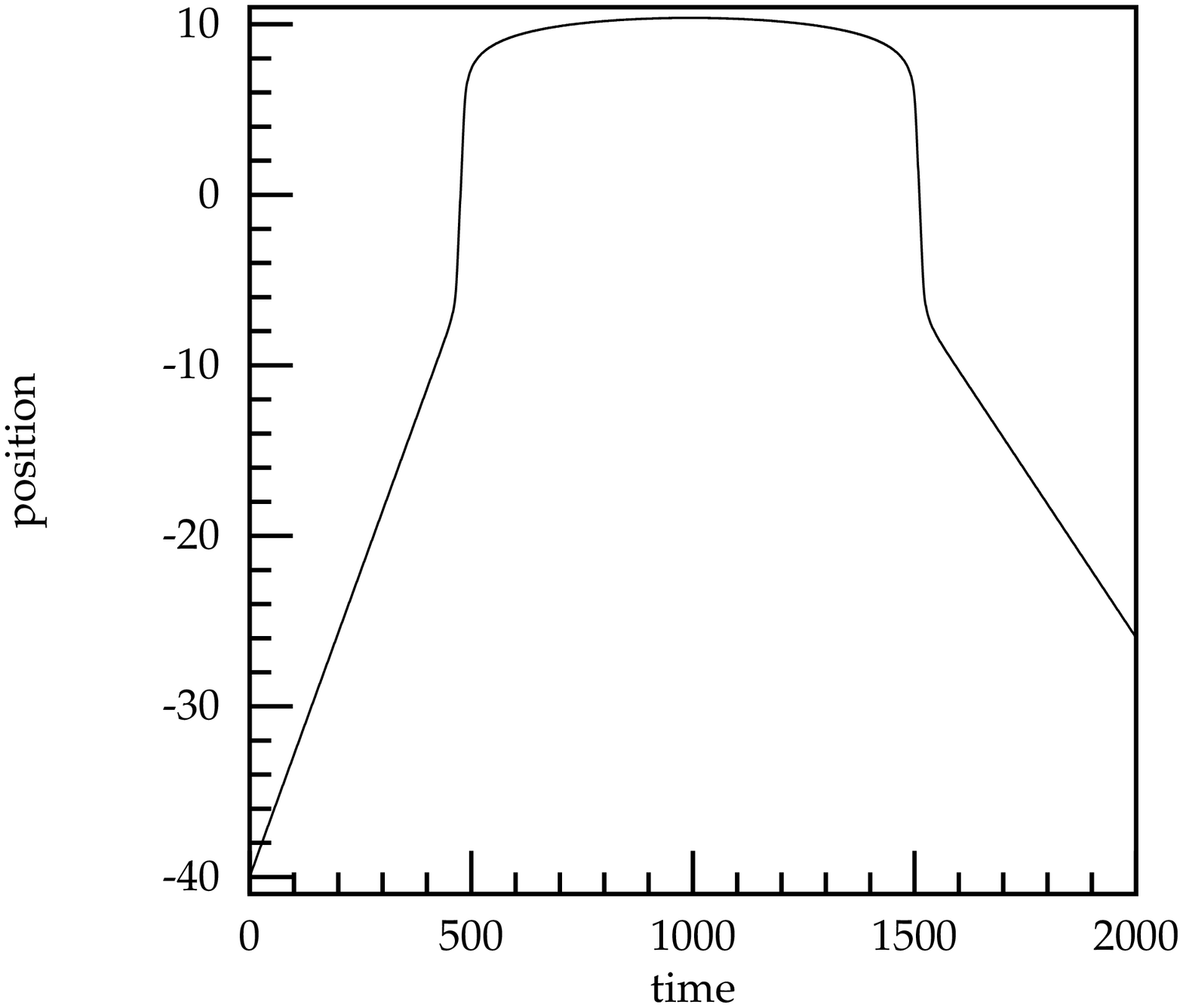}}
\put(4,8){a}
\put(12,8){b}
\put(8,0){c}
\end{picture}
\caption{\label{vc_a-0.8k} Typical trajectories ($p=5$, $\lambda=0.8$).  a) 
Below 'threshold' ($v=0.071535$), b) Transmitted ($v=0.077983$), 
c) Reflected  ($v=0.071514$).
}
\end{figure}

Of course - the reflected trajectory is somewhat unexpected; this is what one 
would expect in a quantum system but here we have a completely classical system
and we have a reflection by the well. Clearly, this reflection must be somewhat
related to the interaction of the deformed kink  with the radiation that 
is present in the well.
To look at this in more detail we have looked also at a very narrow well.

\begin{figure}[htbp]
\unitlength1cm \hfil
\begin{picture}(16,8)
 \epsfxsize=8cm \put(0,0){\epsffile{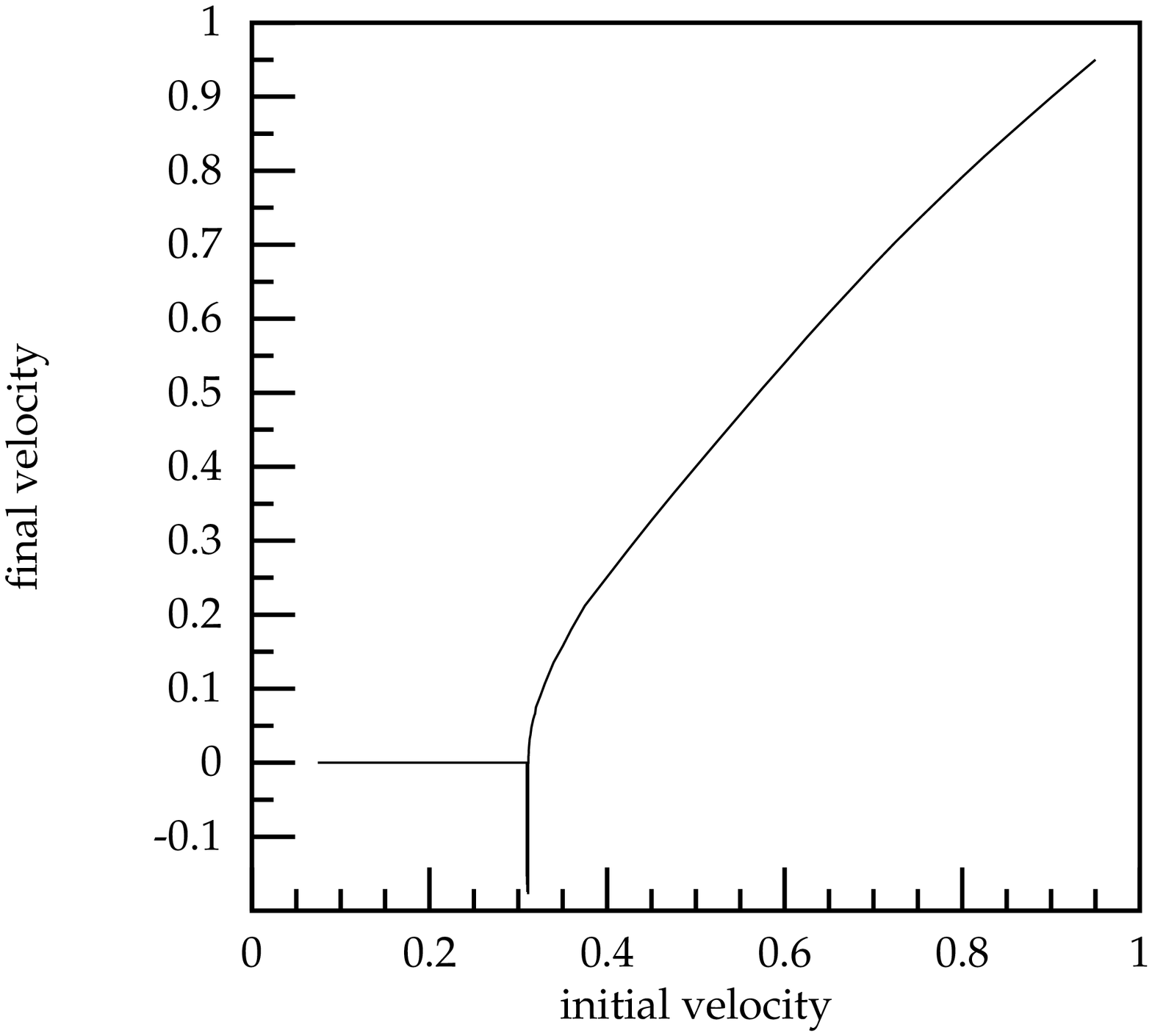}}
 \epsfxsize=8cm \put(8,0){\epsffile{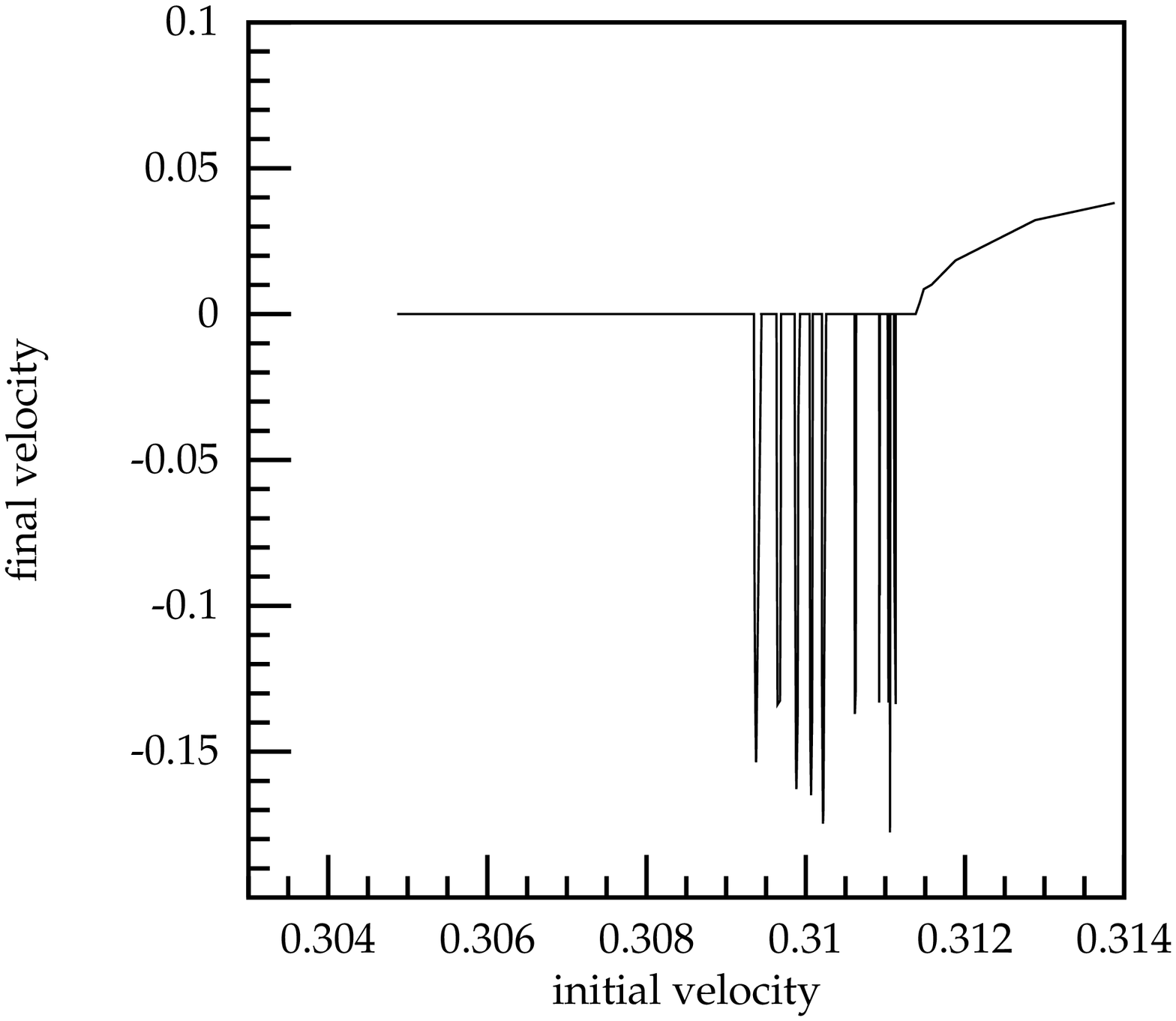}}
\put(4,0){a}
\put(12,0){b}
\end{picture}
\caption{\label{vout_L10k2} Velocity of the kink leaving the well as a 
function of  
its velocity as it approaches the well ($p=0.5$, $\lambda_0=0.2$). 
a) Full plot. b) A close up for small velocities.}
\end{figure}

\subsection{Very Narrow Well}
This time we took $p=0.5$ and $\lambda_0=0.2$. The obtained curve of the 
velocities is shown in fig. 4a.
Again we see an interesting behaviour close to the threshold - and blowing 
it up we get fig. 4b.

Are these results just numerical artifacts? To answer this question we have 
carried out several tests. First we changed the lattice size (increased the 
number of points, changed
the lattice spacing) and changed the time step in the Runge Kutta scheme.
However, the observed pattern of reflections was reproduced in all simulations.
We have also varied the discretisation scheme and, among others, considered 
also the `topological discretisation' of Speight and Ward \cite{ward}.
All simulations using these discretisations exibited similar patterns, with 
the values of the velocities
esentially unchanged (the values changed in the fourth decimal points).
Hence we believe the effect to be genuine and so we are left with having to 
explain its origin.

Performing a literature search we have found the above mentioned paper by  Fei 
et al \cite{older}.
In that paper the authors  studied the scattering of Sine-Gordon kinks on a one
lattice point impurity (at one lattice site the potential $\lambda \varphi^2$ 
was changed
to  a different value - $\lambda' \varphi^2$). Fig. 11 of that paper looks 
amazingly similar
to our fig. 4b. Of course, our potential corresponds to a superposition of 
defects of Fei et al so not surprisingly the pattern is more evident in a 
system involving a smaller well. However, the fact that the effect persists 
and is seen for larger wells
suggests that the phenomenon is more fundamental in nature.

The phenomenon observed by Fei et al was recently explained by Goodwin et al 
\cite{Goodman}
in terms of a two mode model invoving the interaction of the kink with the 
radiation at the
impurity point. Clearly in our case we have two special points (at the two 
'edges' of the well
where the potential changes). Hence it may make sense to develop a model 
based on osciallations at these
two points. This is what we do in the next section.

\section{Our effective model}

Following Goodman et al \cite{Goodman} we consider the following ansatz
\begin{equation}
\varphi(x,t)\,=\,2\,\tan^{-1}\,\exp\left(\theta(t)(x-x_0(t))\right)\,+
\,a(t)g(x)\,+\,b(t)h(x),
\label{ansatz}
\end{equation}
where we take $g(x)$ and $h(x)$ in the form:
\begin{equation}
g(x)\,=\,\exp\left(-\frac{\vert x-p\vert}{2}D\right),
\qquad h(x)\,=\,\exp\left(-\frac{\vert x-p\vert}{2}F\right),
\end{equation}
and 
\begin{equation}
D\,=\,\begin{cases}A,\quad x>p\cr B,\quad x<p\cr\end{cases},\qquad F\,=
\,\begin{cases}B,\quad x>-p\cr A,\quad x<-p\cr\end{cases}.
\end{equation}
Thus we have allowed the kink to change its position $x_0(t)$ and its slope
$\theta(t)$, and $a(t)$ and $b(t)$ represent the excitations of the vacuum.

To obtain an effective model we put this expression for $\varphi(x,t)$ into 
the Lagrangian density (\ref{lagrangians})
and attempted to integrate it over $x$. However, as this leads to rather 
complicated expressions which we had not succeeded to calculate analytically 
we resorted to an expansion
(in the peturbation due to the well).
Hence we put $\varphi=\varphi_0+\delta \varphi$, where
\begin{equation}
\varphi_0 \,=\,2\,\tan^{-1}\,\exp\left(\theta(t)(x-x_0(t))\right)
\end{equation}
and
\begin{equation}
\delta \varphi\,=\,\,a(t)g(x)\,+\,b(t)h(x).
\end{equation}
Next we expanded:
\begin{equation}
\sin^2(\varphi_0+\delta \varphi)\,\sim\,\sin(\varphi_0)\,+\,\sin(2\varphi_0)\,
\delta \varphi
\,+\,\cos(2\varphi)\,(\delta\varphi)^2\,...
\end{equation}

Then as 
\begin{equation}
\frac{\partial\varphi}{\partial t}\,=\,2\cos^2(\frac{\varphi}{2})\left[
\dot \theta (x-x_0)\,-\,\theta \dot x_0\right]\,e^{\theta(x-x_0)}\,+\, \dot a 
\,g(x)\,+\,
\dot b \,h(x)
\end{equation} 
we found that
\begin{equation}
\label{timetheta}
T_1\,=\,\int_{-\infty}^{\infty}\,dx (\frac{\partial\varphi_0}{\partial t})^2\,=\,
\frac{1}{12}\frac{\pi^2 \dot \theta^2}{\theta^3}\,+\,\theta \dot x_0^2
\end{equation}
The $\delta \varphi$ terms give us by themselves
\begin{equation}
\label{timeandb}
T_2\,=\,\frac{1}{2}(\dot a^2\,+\,\dot b^2)\left(\frac{1}{B}+\frac{1}{A}\right)
\,+\,\dot a\dot b
\left[\frac{4}{A+B}e^{-Bp}\,+\,\frac{2}{A-B}\left(e^{-pB}-e^{-pA}\right)\right].
\end{equation}
Finally, we calculated the `crossed terms' and we got (for the terms involving 
$\dot x_0$ and $\dot a$ and $\dot b$)
\begin{equation}
\label{crossedone}
T_3\,=\,-2 \theta \dot x_0\,\dot a\,\frac{K}{1+K^2}\,\left[\frac{1}{\theta +
\frac{B}{2}}\,+\,\frac{1}{\frac{A}{2}-\theta}\right],
\end{equation}
where $K=\exp(\theta(p-x_0)$ and a similar expression for $-2 \theta 
\dot x_0\dot b$, except
that this $K$ was replaced by \hfill \break
 $\tilde K=\exp(\theta(-p-x_0))$.
For the terms involving $\dot \theta$ and $\dot a$ we got
\begin{equation}
\label{crossedtwo}
T_4\,=\,2\dot \theta \,\dot a\,\frac{K}{1+K^2}\left\{\left[\frac{1}{(\frac{A}{2}
-\theta)^2} - \frac{1}{(\frac{B}{2}+\theta)^2}\right]
\,+\,\frac{p(1-K^2)}{(1+K^2)}\left[\frac{1}{\theta +\frac{B}{2}}
+\frac{1}{\frac{A}{2}-\theta}\right]\right\}
\end{equation}
The expression involving $\dot \theta \,\dot b$ was again the same with $K$ 
replaced by $\tilde K$.
These expressions are not exact; we obtained them by making several 
approximations of the type 
\begin{equation}
K\,\int_0^{\infty}\,dz\,{e^{-({\theta}+{\frac{B}{2}})z}\over
1+K^2e^{-2\theta z}}\,\sim \,\frac{K}{1+K^2}\,\frac{1}{\left(\theta +
\frac{B}{2}\right)},
\end{equation}
which should be reasonably reliable given the exponential form of the
dependence of the integrands on $z$ {\it etc}.

Next we calculated the contribution due to 
$({\partial \varphi\over \partial x})^2$.
Performing the integrations, and making similar approximations as before we got
\begin{equation}
\label{derxa}
P_1\,=\,\theta\,+\,\frac{(a^2+b^2)}{8}\,+\,\frac{ab}{4}\left[4(A+B)e^{-Bp}+
\frac{(A+B)^2}{2(A-B)}
\left(e^{-pB}-e^{-pA}\right)\right]
\end{equation}
plus terms linear in $a$ and $b$.
They are given by
\begin{equation}
\label{derxb}
P_2\,=\,-\frac{A+B}{2}\left(\frac{1}{\frac{A}{2}-\theta}+\frac{1}{\frac{B}{2}
+\theta}\right)\left(\theta a 
\frac{K}{1+K^2}\,+\,\theta b \frac{\tilde K}{1+\tilde K^2}\right).
\end{equation}

Finally we had to add the $\lambda^2\sin^2(\varphi)$ terms.
The contribution due to the well is given by
\begin{equation}
\label{well}
\lambda_0^2 \,\int_{-p}^p\,\sin^2(\varphi_0)\,dx \, =\,
\lambda_0^2 \,\int_{-p}^p \frac{e^{2\theta(x-x_0)}}{[1+e^{2\theta(x-x_0)}]^2}\,dx
\end{equation}
$$=\,{\lambda_0^2\over \theta} {\sinh(2\theta p)\over (\cosh(2\theta x_0)
+\cosh(2\theta p)}.$$

The contribution due to $\int_{-\infty}^{\infty}\,dx\,\lambda_0^2\sin^2(\varphi)$
can be calculated in a similar way and we get
\begin{equation}
P_3\,=\,\frac{\lambda^2_1}{\theta}
\end{equation}
from the $\varphi_0$ term, 
\begin{equation}
P_4\,=\,2\lambda^2_1\, \left(\frac{1}{\frac{A}{2}-\theta}+\frac{1}{\frac{B}{2}
+\theta}\right)
\left[ a\frac{K(1-K^2)}{(1+K^2)^2}\,+\,b\frac{\tilde K(1-\tilde K^2)}{(1+\tilde 
K^2)^2}\right]
\end{equation}
from the terms linear in $\delta \varphi$ and
\begin{equation}
P_5\,=\,\frac{\lambda_1^2}{2}\left[(a^2+b^2)\left(\frac{1}{B}+
\frac{1}{A}\right)\,+\,ab\left(\frac{4}{A+B}e^{-Bp}
\,+\,\frac{2}{A-B}\left(e^{-pB}-e^{-pA}\right)\right)\right].
\end{equation}

Adding all these terms together we get the Lagrangian for our effective model.
This Lagrangian is given by
$$ L\,=\,\frac{\pi^2}{12}\frac{\dot \theta^2}{\theta^3}\,+\,\theta\dot x^2\,+
\frac{1}{2}(\dot a^2+\dot b^2)B_0
\,+\,\dot a \dot bB_1\,-\,2\theta \dot x_0(\dot aD_0(\theta)+\dot bD_1(\theta))$$
\begin{equation}
\label{newl}
+
2\dot \theta (\dot aD_2(\theta)+\dot bD_3(\theta))\,-\,\theta \,-\,
\frac{(a^2+b^2)}{8}(A+B)
\,-\,\frac{ab}{4}\,+\,\frac{A+B}{2}\theta(aD_0(\theta)+bD_1(\theta))
\end{equation}
$$-2\epsilon\lambda_1^2D_6(\theta,x_0)\,-\,\frac{\lambda_1^2}{\theta}\,
-2\lambda_1^2(aD_4(\theta)
+bD_5(\theta))\,-\,\lambda_1^2\frac{ab}{2}B_1\,-\,
\lambda_1^2\frac{(a^2+b^2)}{2}B_0,
$$
where 
$$ B_0\,=\, {1\over B}\,+\,{1\over A},\,\quad\qquad B_1\,=\,
{4\over A+B}e^{-pB}+{2\over A-B}\left(e^{-pB}-e^{-pA}\right)$$
and where the 7 functions $D_i$ $i=0,..6$ are given by:

$$
D_0(\theta)\,=\,{K\over 1+K^2}\left[{1\over \theta +0.5B}\,+\,
{1\over 0.5A-\theta}\right],
$$
$$D_2(\theta)\,=\,{K\over 1+K^2}\left[-{1\over (\theta +0.5B)^2})\,+\,
{1\over (0.5A-\theta)^2}\right]\,+\,p{1-K^2\over 1+K^2}
\left[{1\over \theta +0.5B}\,+\,{1\over 0.5A-\theta}\right],$$
$$ D_4(\theta)\,=\,D_0(\theta)\,{1-K^2\over 1+K^2}.$$
The functions $D_1(\theta)$ and $D_5(\theta)$ have the same form 
as $D_0(\theta)$ and $D_4(\theta)$, respectively, after the replacement 
$K\rightarrow \tilde K$ and
$$D_3(\theta)\,=\,{\tilde K\over 1+\tilde K^2}
\left[-{1\over (\theta +0.5B)^2})\,+\,{1\over (0.5A-\theta)^2}\right]\,
-\,p{1-\tilde K^2\over 1+\tilde K^2}\left[{1\over \theta +0.5B}\,+
\,{1\over 0.5A-\theta}\right].$$

Finally $D_6(\theta,x_0)$ is given by
$$
D_6(\theta,x_0)={1\over \theta}{\sinh(2\theta p)\over \cosh(\theta(x_0+p))\,
\cosh(\theta(p-x_0))}.$$

The derived Lagrangian involves 4 variables $x_0,\theta,a$ and $b$ but its 
equations are rather complicated.
Hence we have solved them numerically.
In the next section we discuss our solutions of the equations which follow 
from the Lagrangian (\ref{newl}).

\section{Results in our effective model}

We have looked at the equations for $\theta$, $x_0$ and $a$ and $b$, which 
follow from 
the Lagrangian (\ref{newl}) and solved them numerically.
As our initial conditions we took $x_0=-40$, $\theta=1$ and set, initially, 
$a=b=0$. Of course, we also put $\dot a=\dot b=\dot \theta=0$ and studied the 
behiour of our system
as a function of $\dot x_0$. We also varied a little the parameters $A$ and 
$B$, which
appear in the description of the effects due to the well. For most of our 
work we used
the values around 0.5 (and used the fact that we expect $A^2\sim \epsilon +B^2$.
To have the well similar to the one we used in the full simulation (shallow 
well)
we put $\epsilon \sim 0.16$. This is due to the fact that the linearised 
equations ({\it ie} for the waves) differ, inside and outside the well,
by a term proportional to $(\lambda_0^2-1)u$. In our case this translates
to $A^2-B^2\sim \epsilon$ and so for a shallow well we have 
$\epsilon\sim 1-\lambda_0^2=0.16$.
We simulated the time
evolution using the 4th order Runge Kutta method and have found that the well 
distorts the 
kink quite strongly but, as expected, it can trap the kink like in the 
original 
Sine-Gordon model. However, as the effective system does not absorb energy 
after a few 
bounces the kink can escape (forwards or backwards). 
Of course in the real system there are many degrees of freedom of radiation
which can get generated in the hole and such bounces are very rare.
So to model these `extra' modes of radiation which take the energy
out of the modes we are describing we introduced 
an absorption of the oscillations of $\theta(t)$, {\it ie.} we added a term 
proportional to 
$\dot \theta$ in the equation for $\theta$. This has, indeed, reduced the 
oscillations in $\theta$ and made the model more realistic.

In fig. 5 we present the curve of $v_{out}$ as a function of $v_{in}$ obtained 
in our model, {\it ie} based on the Lagrangian (\ref{newl}).
\begin{figure}
\begin{center}
\includegraphics[width=10cm]{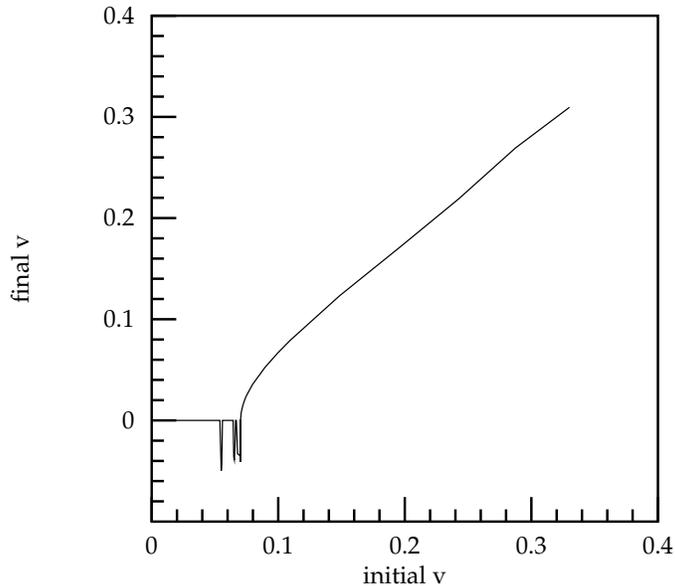}
\caption{Outgoing velocities as a function of incoming ones - in our 
effective model.}
\end{center}
\end{figure}
We note that although our effective model is quite crude it reproduces fig. 3b 
rather well.
Hence we believe that the mechanism of Goodman et al \cite{Goodman} explains 
the behaviour 
of the kinks in our case too. Of course, we could check this in more detail by 
making fewer
approximations and, perhaps, performing  the evaluation of all the difficult 
integrals numerically, but we are not sure that the extra effort required would
be justified.

\section{Another effective Model}

Given that our effective model reproduces the results of our 
simulations rather well we have decided to check the generality of this
observation - is this related to the existence of standing 
waves in tbe neighbourhood of the well or is our model somewhat unique.
Hence we considered another model this time involving a standing wave in the 
well. The idea here is to explore further whether the relections observed in 
the full model can be related to the interaction 
of the soliton with the waves in the well.
Hence this time we have taken for our field
configuration
 $\varphi=\varphi_0+\delta \varphi$, where, as before,
\begin{equation}
\varphi_0 \,=\,2\,\tan^{-1}\,\exp\left(\theta(t)(x-x_0(t))\right)
\end{equation}
and
\begin{equation}
\delta \varphi\,=\,\,a(t)g(x).
\end{equation}
This time for $g(x)$ we have taken
\begin{equation}
g(x)\,=\,\cos({\pi x\over 2p})\,-\,\frac{1}{3}\cos({3\pi x\over 2p})
\end{equation}
inside the well ({\it ie} for $|x|<p$ and zero otherwise).
Our $g(x)$, and its derivative, vanish at $x=\pm p$
and the idea is that as the soliton enters the well its slope  $\theta(t)$ 
changes. This excites
the modes described by $g(x)$ and so $a(t)$ becomes nonzero. This puts some 
energy into these modes which then interact with the soliton.
 Due to this interaction this
energy can, every now and 
then, be given 
back to the soliton resulting in its reflection or transmission.

Like in the previous model, we have put the expression for $\varphi$ into 
the original Lagrangian, integrated over $x$ obtaining an effective model
involving $x_0$, $\theta$ and $a$. Then we performed some simulations
starting with a soliton originally far from the well moving with some velocity
towards it. Like in the previous model we have absorbed energy through a term
propertional to ${\partial \theta\over \partial t}$ in the equation 
for $\theta$. Also, like in the previous case we have found some reflections
below the threshold for the scattering through the well.
Their details depend a little on the strength of the absorption; at no 
absorption 
we have reflections, and transmissions forward, at much lower velocities, 
the higher absorptions kill the reflections and transmissions. They also 
raise the threshold for transmission. 
Of course, it is difficult to estimate reliably the degree of absorption
as in the full model there are many modes of oscillations in the well.
So in the end we have used very low absorption; then the threshold for 
transmission was close enough to what is seen in the full simulation
but we had more reflections below threshold. To improve the model 
we should have used more modes in the well but this would have resulted
in a more complicated model with more degrees of freedom. As we have 
only wanted to test our basic idea (that the reflections are due to the 
interaction
of the soliton with radiation modes), in this paper, we decided
to look at a model which is as simple as possible {\it ie} with a very few
degrees of freedom.
Our choice of two low lying modes was dictated by simplicity and the basic 
belief that the lower modes would get excited first and so are more important.
In fig. 6 we present the curve of $v_{out}$ as a function of $v_{in}$ obtained 
in this (second) model (with a reasonble absorption).
\begin{figure}
\begin{center}
\includegraphics[width=10cm]{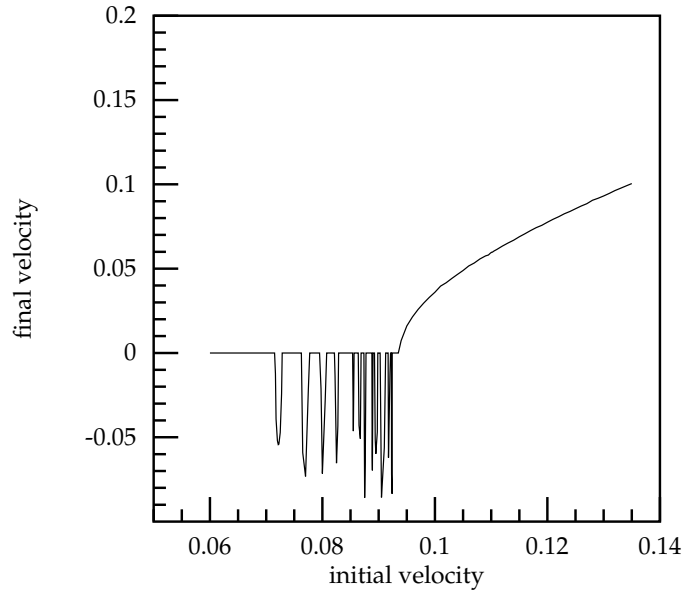}
\caption{Outgoing velocities as a function of incoming ones - in our 
second effective model.}
\end{center}
\end{figure}
Once again, we see that the model reproduces  well the pattern of the 
simulations of the full Sine-Gordon model (fig. 3b).
\begin{figure}
\begin{center}
\includegraphics[width=10cm]{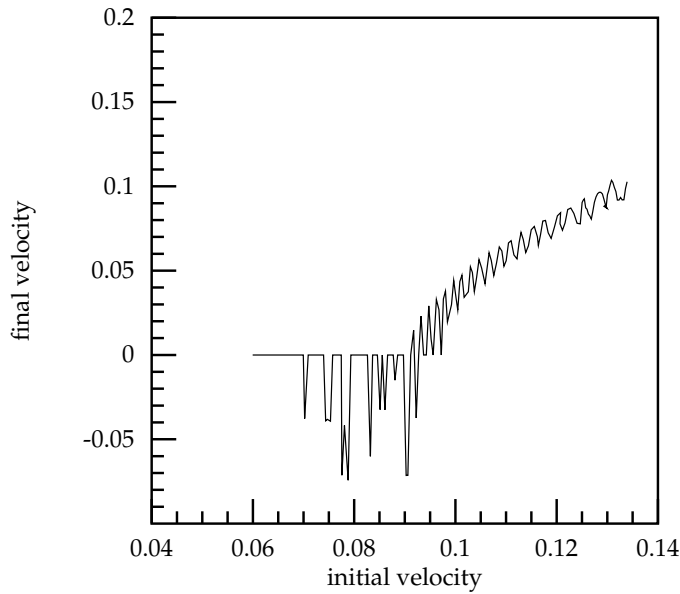}
\caption{Outgoing velocities as a function of incoming ones - in our second 
effective model without relativistic corrections.}
\end{center}
\end{figure}

\section{Conclusions}

We have looked at a system involving a Sine-Gordon kink scattering
 on a `well'-like potential.

We have found that, like in (2+1) dimensions,
when the kink was sent towards the well it gained some energy 
as it entered the well. Some of this
energy was converted into kinetic energy of the kink, some was
radiated away.  So when the kink tried to `get out' of the well it
had less kinetic energy than at its entry and, when this energy was
too low it remained trapped in the well. However, as the soliton moved
in the `well' it interacted with the radiation in the well and at some specific
values of the initial velocity this interaction resulted in the kink
being ejected backwards from the well (with much reduced velocity).
Thus, seen from outside, the well acted as if it reflected the kink, something 
which is seen in quantum systems but which is less well known 
in classical systems.

We have performed many numerical simulations to make sure that the observed
behaviour is not an unexpected artifact of our numerical procedures and 
the pattern survived all applied tests. Hence we believe the effect to be 
genuine.

We have noted that a similar behaviour was observed many years ago by Fei et al
who studied the scattering of kinks on a one-point impurity. This behaviour 
was recently explained by Goodman et al \cite{Goodman} as a two-bounce 
resonance between the 
kink and the oscillation of the defect. This has made us to consider two models
of a similar nature. Both models are very simple, clearly too simple, but
we wondered whether they would qualitatively reproduce the observed effects.
Both involve kinks interacting with radiation and 
we generate this by taking an ansatz involving
a kink (which can vary its position and slope) plus a couple of radiation 
modes.
This ansatz is then put into the full equation which are then integrated
out resulting in a model involving a few variables; namely, the position
and the slope of the kink and the coefficients of the oscillation modes
of the vacuum (modeling radiation).

The first model involved taking standing waves that are located at the eges
of the well; the other one involved just one wave in the well.
We have found that both models reproduced the main features of the observed 
behaviour  quite well
suggesting that the mechanism of Goodman et al is more general in nature and 
that, in general, the
observed phenomenon of reflection of the solitons on the well is related
to their interaction with the waves in the well.
\begin{figure}
\begin{center}
\includegraphics[width=10cm]{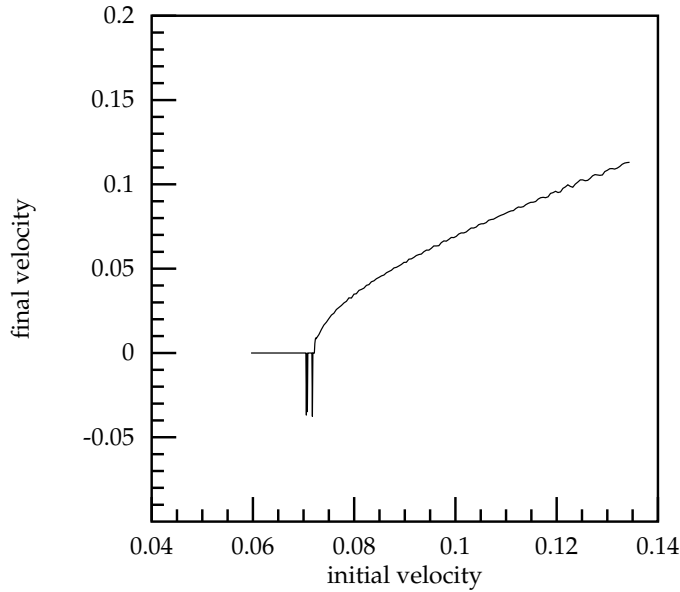}
\caption{Outgoing velocities as a function of incoming ones - in the
full simulation without relativistic corrections}
\end{center}
\end{figure}
Of course, both models are too simplistic; to describe properly the full
process we have to understand better which modes of radiation are important
and why. This involves more work and is planned for the future. 
However, the work done so far suggests that we are on the right track
and gives us encouragement for the further study. This is confirmed further
by what we saw in a two-dimensional model \cite{two} in which the solitons
have genuine vibrational as well as radiation modes. 

Incidentally, in our calculation we have used the correct initial condition
{\it ie} with the correct relativistic factors ($\gamma={1\over \sqrt{1-v^2}}$).
Had we ignored them and used 
their norelativistic form ({\it ie} 
not modified $\theta$ in (10) and (26)) we would have obtained instead of 
Fig. 6 
the dependence which is shown in Fig. 7.
We note an interesting oscillatory behaviour. Of course, this oscillation
is overemphasised by the use of too few radiation modes but we wondered
whether it would be seen in the full model too ({\it ie} whether the addition
of further modes would wash them out).
Hence we have redone the full simulations also without the relativistic
factors.
Our nonrelativisitc curve is shown in fig. 8. We note the extra oscillations.
Their origin lies clearly in the fact that the absence of $\gamma$ factors 
induces initial distortions resulting in the change of $\theta$.  This affects
the phase of the soliton and so alters its interaction with the wave
in the well resulting in the curve shown in fig. 8. It is interesting to note 
that our effective models also reproduce these oscillations, thus giving 
further support for the validity of our claim that the interaction with the 
well
proceeds through the generation of standing waves and their intereference with 
the solitonic fields. Of course the agreement between the results of the full
simulations and of the effective models is only qualitative in nature as in 
our effective models we used only some standing waves which were chosen 
somewhat ad hoc. To get a quantitative agreement we have to determine the 
relative
importance of different waves - this problem is currently under consideration.

Several real physical systems are described by the sine Gordon equation,
especially in solid state physics, and it would be interesting to see what
the physical implications of a position dependant potential, like the one
used in this paper, would be.

\section*{Acknowledgements}
This investigation is a natural follow up of the work
on (2+1) dimensional topological solitons originally performed
in collaboration with Joachim Brand. We would like to thank him 
for this collaboration.

{}

\end{document}